\documentstyle[aps,epsf]{revtex}
\def\p{^\prime}
\def\h#1{\hat{#1}}

\def\gam{\gamma}
\def\gamxc{\h\gamma_{\mbox{\tiny XC}}}
\def\gamh{\h\gamma_{\mbox{\tiny H}}}

\def\k{{\bf k}}
\def\kp{{\bf k^\prime}}
\def\x{{{\bf x}_1}}
\def\xx{{{\bf x}_2}}
\def\xp{{{\bf x}^\prime_1}}
\def\xxp{{{\bf x}^\prime_2}}
\def\r{{{\bf r}_1}}
\def\rr{{{\bf r}_2}}

\def\<{\langle}
\def\>{\rangle}
\def\tpc{(2\pi)^3}
\def\tpone{\hbox{\vbox{%
\hbox to 10pt{\hfil$\otimes$\hfil}%
\hrule height 1pt width 0pt%
\hbox to 10pt{\hfil\tiny I\hfil}%
\hrule height -5pt depth 5pt}}}
\def\tptwo{\hbox{\vbox{%
\hbox to 10pt{\hfil$\otimes$\hfil}%
\hrule height 1pt width 0pt%
\hbox to 10pt{\hfil\tiny II\hfil}%
\hrule height -5pt depth 5pt}}}
\def\tpthree{\hbox{\vbox{%
\hbox to 10pt{\hfil$\otimes$\hfil}%
\hrule height 1pt width 0pt%
\hbox to 10pt{\hfil\tiny III\hfil}%
\hrule height -5pt depth 5pt}}}
\def\appge{\hbox{\vbox{%
\hbox to 10pt{\hfil$>$\hfil}%
\hrule height 1pt width 0pt%
\hbox to 10pt{\hfil$\sim$\hfil}%
\hrule height -5pt depth 5pt}}\hskip 4pt}

\begin{document}

\twocolumn


\title{Improved tensor--product expansions for the two--particle
density matrix}

\author{G\'abor Cs\'anyi$^\dag$, Stefan Goedecker$^*$ and
T.A. Arias$^\ddag$\\ $\dag$ Department of Physics, Massachusetts
Institute of Technology, Cambridge, MA 02139\\
$^*$ D\'epartement de recherche fondamentale sur la mati\`ere condens\'ee,\\
SP2M/NM, CEA-Grenoble, 38054 Grenoble cedex~9, France\\
$\ddag$ Laboratory of
Atomic and Solid State Physics, Cornell University, Ithaca, NY 14853\\
}



\abstract{ We present a new density--matrix functional within the
recently introduced framework for tensor--product expansions of the
two--particle density matrix.  It performs well both for the
homogeneous electron gas as well as atoms.  For the homogeneous
electron gas, it performs significantly better than all previous
density-matrix functionals, becoming very accurate for high densities
and outperforming Hartree-Fock at metallic valence electron
densities.  For isolated atoms and ions, it is on a par with
generalized gradient approximations to
density-functional theory.  We also present analytic results for the
correlation energy in the low density limit of the free electron gas
for a broad class of such functionals.  \break \vrule height 15pt
width 0pt PACS 71.10.-w 71.15.Mb, submitted to Physical Review Letters}

\maketitle

The long standing quest for an accurate description of electronic
correlation with effective single--particle theories has renewed
interest in density--matrix functionals in recent years.  The success
of density functional approaches, based on Hohenberg--Kohn--Sham
theory \cite{HohenbergKohn,KohnSham}, is well established
\cite{dftreview}.  Unfortunately, the known approximations to the
requisite universal energy functional of the density are not
sufficiently accurate to predict the rates of chemical and physical
processes at room temperatures.  This theory can      easily be
generalized to show that there is a universal correlation energy
functional of the single--particle density {\em matrix} as
well~\cite{Levy79,Valone80,DonellyParr78,Zumbach85,Davidson,Levy}.
A density matrix functional requires an expression of the two--body matrix
in terms of the one-body density matrix. An ansatz for the two--body matrix
containing one free parameter was proposed by M\"uller~\cite{Mueller}, and
the case where this parameter is set to 1/2 has been
discussed by Buijse~\cite{Buijse}.  In addition, Buijse attempted to perform
self-consistent calculations but optimized only the lowest
two natural orbitals and occupation numbers and only for the
H$_2$ molecule. Independently, Goedecker and Umrigar~\cite{Goedecker}
proposed a functional based on a very similar, but self-interaction corrected, two--body matrix
and derived the Euler--Lagrange formalism necessary for the numerical minimization
of the functional.  Optimizing typically 50 natural orbitals and occupation
numbers, necessary to achieve convergence, for a variety of atoms and ions
it was found that the Goedecker-Umrigar (GU)
functional yielded energies that were comparable or better than those from
the generalized gradient approximation (GGA) in density functional theory
and densities that were better than those from GGA.
Following these first encouraging numerical results
several groups of authors have recently proposed and studied the accuracy of
various new approximations to the density-matrix
functional\cite{Goedecker2,CSG,Cio,Cio2,lopez,hennig,bba1,bba2}.
However, numerical studies~\cite{CSG} demonstrate that
density--matrix functionals, to date, give poor
correlation energies for the homogeneous electron gas.

It has been shown analytically that the {\em per particle} correlation
energy of the homogeneous electron gas in the low density limit approaches a non-zero
constant\cite{Cio} for these initial functionals.
Building on our tensor--product formalism\cite{CSG}, below we extend
this proof and demonstrate this shortcoming for a broad class of
functionals.  As regions of low densities contribute relatively little
to the total energies of solids and molecules, we focus our attention
on improving the results for the high and intermediate density
regimes.  We present a new density--matrix functional which is more
than an order of magnitude more accurate in the high--density limit
for the homogeneous electron gas
than those proposed previously and also represents a significant
improvement in the intermediate regime.  In atoms, the new functional
performs similarly to
the generalized gradient approximation (GGA)\cite{PBE} to density
functional theory but not as well as the GU functional.
Because selfinteraction corrections complicate numerical calculations,
we do not enforce them on our new functional in the present work.

{\em Notation ---} For a system of $N$ electrons, the exact energy
functional of the {\em two}--body density matrix $\gam(\x\xx,\xp\xxp) \equiv
\langle \Psi | \h\psi^\dagger(\xp) \h\psi^\dagger(\xxp) \h\psi(\xx)
\h\psi(\x) | \Psi \rangle/2$,
is
\begin{eqnarray}
E & = & \int d\x \left. \left(\left[-\frac12\nabla^2_\r
+U(\r)\right]
n(\x, \xp)\right)\right|_{\xp=\x} \label{totE}\label{totalenergy}\\
&&+ \int d\x d\xx\, \frac{\gam(\x\xx,\x\xx)}{|\r-\rr|}. \nonumber
\end{eqnarray}
Here and throughout, we work in atomic units, and the ${\bf x}_i$ are
compound coordinates representing both position ${\bf r}_i$ and spin
$s_i$, so that integration over ${\bf x}_i$ represents integration
over space and summation over spin channels.  In the above expression,
the external potential and one--body density matrix are $U({\bf r})$
and $n(\x, \xp) \equiv \langle \Psi | \h\psi^\dagger(\xp) \h\psi(\x) |
\Psi \rangle$, respectively.  Finally, we note that simple integration
of the definitions reveals the sum--rule which connects the one-- and
two--body matrices,
\begin{equation}
\left(\frac{N-1}2\right) n(\x,\xp) = \int\!d\xx\,\gam(\x\xx,\xp\xx).
\label{sumrule}
\end{equation}

As in \cite{CSG}, we expand $\gam$ into Hartree and
exchange--correlation contributions,
\begin{equation}
\h\gam = \gamh + \gamxc\label{gamxcdef}.
\end{equation}
By separation of variables, it is always possible to expand the
unknown exchange--correlation part as a (potentially infinite) sum of
tensor products,
\begin{equation}
\gamxc = \sum_i \,\, \h u_i\otimes\h v_i\label{gamexpansion}.
\end{equation}
In this context we view the two--body density matrix as a two--body
operator (a function of four variables) and $\h u_i$ and $\h v_i$ as
one--body operators (functions of two variables).  Note that throughout we
employ the twisted tensor product (denoted as ``Type III'' product in
\cite{CSG}),
\[
[ u \otimes v] (\x\xp, \xx\xxp) \equiv
u(\x,\xxp) v(\xp,\xx),
\]
to combine the one--body operators.

Remarkably, given such an expansion truncated to finite order, the
combination of the symmetry constraints on $\gam$ with the sum--rule
is sometimes sufficient to determine the full four--variable function
$\gam$ directly in terms of the two--variable density matrix $n$.
Under these conditions, one can then evaluate the energy as an
explicit functional of the density matrix using (\ref{totalenergy}).
Minimizing the density--matrix functional over all physical density
matrices then yields the ground--state of the system under this
approximation.

{\em New Functional ---} Considerable freedom remains in the
terms for the expansion (\ref{gamexpansion}).  Previously, we explored
two choices, ``Corrected Hartree--Fock'' (CHF)\cite{CSG}, and the
``Corrected Hartree'' (CH) functional,
\begin{eqnarray*}
\gamxc^{\mbox{\tiny CHF}} &=& - \frac12(\h n\otimes \h n) - \frac12(\sqrt{\h n(1-\h n)}\otimes \sqrt{\h n(1-\h n)}),\\
\gamxc^{\mbox{\tiny CH}} &=& - \frac12(\sqrt{\h n}\otimes \sqrt{\h n}),
\end{eqnarray*}
respectively.  The latter is identical to the one proposed by Buijse~\cite{Buijse}
and a special case of the one proposed by M\"{u}ller~\cite{Mueller}.
The GU functional is the CH functional
with a selfinteraction correction. In the case of the homogeneous electron gas
at high densities the GU and CH functionals become identical since the selfinteraction
correction vanishes but at low densities they could differ if the natural orbitals
are localized.
One way to view these functionals is in terms of the
coefficient of the Fock term $\h n\otimes \h n$, which is $-1/2$ in
the CHF functional and is zero in the
CH functional.  Under truncation of the expansion to
one additional term, the value of this coefficient, through the
sum--rule, then uniquely determines the form of the remaining tensor
product.

\begin{figure}
\epsfxsize =3in
\epsffile{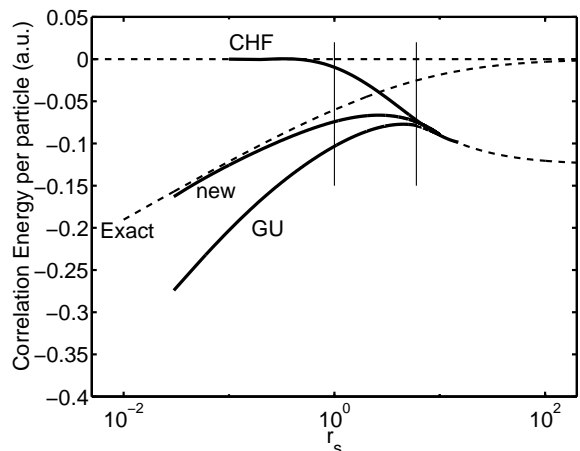}
\caption{Correlation energies of the three functionals in the text and
the exact result \protect\cite{Mahan,CPA,GMB}, versus the mean
particle separation.  The solid lines are numerical calculations, the
dashed lines are exact results.  The vertical lines indicate the range
of typical metallic valence densities.}
\label{Fig:ener}
\end{figure}

{\em Homogeneous electron gas ---} Figure~\ref{Fig:ener} shows the
correlation energies of the resulting functionals for the
spin--unpolarized homogeneous electron gas.  Corrected Hartree--Fock
leads to no improvement in the correlation energy over Hartree--Fock,
and approaches zero for high electron density.  The CH
functional is a little better, and represents an improvement by a
factor of two over Hartree--Fock at high densities, where it is
over-correlated. Both functionals are over-correlated at low
densities, where they perform badly, with the per particle correlation
energy approaching a {\em non--zero} constant.  This property of the
CH functional has been demonstrated analytically in
\cite{Cio}.

Having seen the behaviour of the above two functionals in the high
density regime, it is natural to look for a form with a Fock
coefficient intermediate to those above.  A
coefficient of $-1/4$ for the Fock term results in the following functional,
\begin{eqnarray}
\gamxc =  -\frac14\left(\h n \otimes \h n + \sqrt{\h n(2-\h n)}
\otimes \sqrt{\h n(2-\h n)}\right).
\end{eqnarray}
Figure \ref{Fig:ener} shows that this new functional results in a
dramatic improvement at high densities (low $r_s$) and significant
improvement at metallic valence densities ($1 < r_s < 6$).  At low
density, the behaviour of {\em all three} functionals becomes the same,
resulting in unrealistically large correlation energies {\em per
particle}.  Note that because this happens at low density, the
contribution of these regions to the total energy is likely to be
small in an inhomogeneous system.
We now explore the reasons for this behaviour to ascertain the
feasibility of making improvements in this regime as well.

\begin{figure}
\epsfxsize = 3in
\epsffile{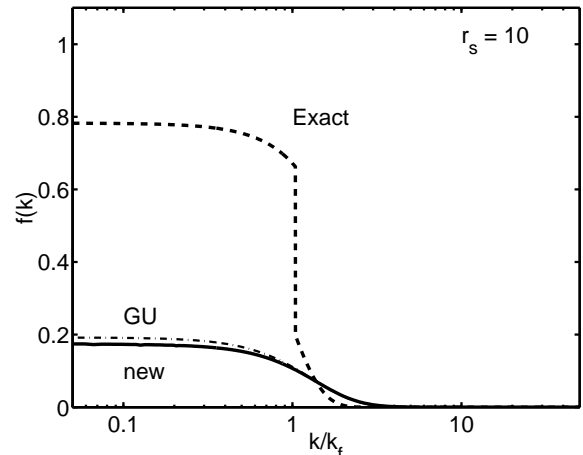}
\caption{The momentum distribution $f(k)$ as a function of the wave
vector, $k$: exact \protect\cite{Farid} (dashed line),
CH functional \protect\cite{Cio} (dash--dotted line) and our
new functional (solid line) at $r_s = 10$.  The Fermi wave vector is
denoted $k_f$}
\label{Fig:mom}
\end{figure}

{\em Low density limit ---} To understand the approach of the three functionals
shown in Fig.~\ref{Fig:ener}
to one another in the low density limit, we recall that
Cioslowski and Pernal\cite{Cio} have determined the predictions of the
CH functional analytically for $r_s$ \appge $5.769$.
Their analysis applies, provided that $\gamxc$ is represented as a
single tensor product of a specific homogeneous form.  As an
illustration of the anomalous correlation behaviour which these
functionals exhibit in the low density limit, Figure~\ref{Fig:mom}
compares the known many--body momentum distribution with numerical
results for our new functional and the analytic distribution from
\cite{Cio} derived for the CH functional for $r_s =
10$.  Both density--matrix functionals have very low occupation
numbers, and lose the characteristic signature of the sharp drop near
the Fermi wave vector evident in the exact curve.

We now show that, as a consequence of the sum--rule, all of the above
functionals reduce to this particular form in the limit of high $r_s$.
Fourier transforming (\ref{totalenergy}), the total energy in the
homogeneous electron gas becomes
\begin{equation}
E = 2\int\!{Vd\k\over\tpc}{{\k^2\over2}f_\k} - {1\over V}\int\!{V^2 d\k d\kp \over (2\pi)^6} g_{\k \kp}
{4\pi\over|\k\p-\k|^2}, \label{efunctional}
\end{equation}
where $g_{\k \kp}$ is the momentum space representation of $\gamxc$.
In the case that $\gamxc$ is expanded as a sum of tensor products of
single particle operators, its momentum space representation will be a
sum of tensor products of the momentum space representation of those
operators.  Because our one--body operators are explicit functions of
the density matrix, which for the homogeneous electron gas is diagonal
in the momentum representation, their momentum space representations
are also diagonal.  Thus, their tensor products become simple
products, and we have
\begin{equation} \label{gkk}
g_{\k \kp} = \Omega_1[n(\k), n(\kp)] + \Omega_2[n(\k), n(\kp)] + \ldots,
\end{equation}
where the $\Omega$ are the same functions of the density matrix which
appear in the tensor products expressing $\gamxc$ but with the
tensor--product operator replaced with simple multiplication.  For
example, for the CH functional $g_{\k \kp} = -\frac12
\sqrt{n(\k) n(\kp)}$.

We now show that, in the low--density limit, the analytic solution
presented in \cite{Cio} for the case when $g_{\k \kp}$ consists of a
{\em single} $\Omega$ function of the homogeneous form
\begin{equation}
\Omega[\lambda n(\k), \lambda n(\k)]=
\lambda^{\beta}\Omega[n(\k), n(\kp)],
\label{homo}
\end{equation}
applies more generally to cases with multiple tensor products.  The
analysis of \cite{Cio} begins with the scaling {\em ansatz}
\[
n(\k) = \rho^\Delta\eta(\rho^\nu\k),
\]
where $\rho$ is the total electron density per spin channel, $\nu$ is
a constant, $\eta$ is a universal function that does not depend on the
density explicitly, and $\Delta = 1-3\nu$ due to the constraint of the
total particle density.  Note that as $\rho \rightarrow 0$, $n(\k)
\rightarrow 0$ uniformly (provided that $\Delta > 0$).  This
behaviour, evident in Figure \ref{Fig:mom}, forms the crux of our
argument; we verify it self--consistently below.

In the above limit, all of the $\Omega_i$ in the above functionals
approach homogeneous functions with their own exponents $\beta_i$.
Substituting $n(\k)$ into (\ref{gkk}), we then find that a single
homogeneous function dominates $g_{\k \kp}$, the one with the smallest
$\beta_i$ ($\equiv \beta_{\rm min}$).  Therefore, all the conditions
for the analysis of \cite{Cio} apply, in particular that
\begin{eqnarray*}
\nu &=& \frac{1-\beta_{\rm min}}{3\beta_{\rm min}-2}.\\
\end{eqnarray*}

One can show that not only the above functionals, but any
functional satisfying the sum--rule and dominated by a {\em single}
homogeneous term in the low--density limit, will exhibit $\beta_{\rm
min} = 1$ in that limit.  If indeed a single tensor product term
dominates $\h\gam$, then the symmetry arguments presented in
\cite{CSG} imply that $\gamxc$ must take the form
\begin{equation}
\gamxc = \h u[\h n]\otimes\h u[\h n].
\label{gamxcsingle}
\end{equation}
The sum--rule (\ref{sumrule}) then implies
\begin{equation}
\int\! d\xxp \gamxc = \h n,
\end{equation}
which, using (\ref{gamxcsingle}), can be written as an operator equation
\begin{equation}
(u[\hat n])^2 = \hat n,
\label{u2}
\end{equation}
which implies that $u[\ ]$ must be a homogeneous function with
exponent $1/2$ and therefore that $\gamxc$ is dominated by a term of
homogeneous exponent $\beta_{\rm min} = 1$.  Therefore $\nu = 0$ and
$\Delta = 1$; the fact that $\Delta > 0$ confirms the
self--consistency and the validity of the above analysis.

Ref.~\cite{Cio} analyzes the case $\beta=1$ in detail, finding for
the correlation energy per particle the result
\begin{eqnarray*}
\epsilon_{\tiny c} &\rightarrow -\frac{1}{8}, \\
\end{eqnarray*}
as $\rho \rightarrow 0$.  Figure~\ref{Fig:ener} verifies that, in the
low density limit, indeed the correlation energy of all three
functionals approach one another and, in particular, that they all
approach a non--zero constant correlation energy of $-1/8$ Hartree per
particle.

{\em Atoms ---} The electrons in solids exhibit both free--electron
like and localized atomic--orbital behaviour.  Thus, another
important limiting case for assessing new functionals is their
behavior in atoms.  Table~\ref{table:atoms} presents energies obtained
from the new functional for light atoms and
ions.  For comparison, we include energies obtained from the CH and GU
functionals as well as from the local density approximation (LDA), and
the Perdew--Burke--Ernzerhof (PBE)~\cite{PBE} generalized gradient
approximation.

The table reveals that, on the whole, the new functional performs for
atoms quite comparably to both the CH functional and
to PBE.  In particular, for the two--electron series, the new
functional is somewhat worse, but for the four--electron series, it is
somewhat better.
All density--matrix functionals are generally much
better than LDA.  For the neutral atoms, CH and the present functional
are not quite as good as PBE, but, for the ions both
are better or in some cases similar to PBE.
Note that the CH functional always overestimates the correlation energy,
which is not the case for the new functional.
In the GU functional this overcorrelation
is corrected by the selfinteraction correction and this functional
yields the best energies for atoms and ions.
Thus, we feel that the density--matrix functionals provide an adequate
description of atom--like systems.

{\em Conclusions ---} We have introduced a new density--matrix
functional, which represents a major improvement over existing
functionals of the density matrix.
It is very accurate in the high
density regime of the homogeneous electron gas, significantly improves
the correlation energies at typical valence densities, and it is
comparable
to the generalized gradient approximation in atoms.  Although it is
not as accurate as the GU functional for atoms, we expect it to perform
better than the GU functional for solids.  We have also shown
that, in our tensor--product expansion of the two--body density
matrix, little further improvement can be expected at the present
truncation of two terms.  Our test systems span the range of
environments encountered in solids and molecules, so we conclude that
this new functional is a good candidate to be used in electronic
structure calculations of condensed matter.

We thank Cyrus Umrigar for interesting discussions and helpful comments 
on the manuscript.

\vfill
\break

\begin{table}
\def\tablestrut{\vrule height 10pt depth 7pt width 0pt}
\def\vertline{\vrule width 0.2pt}
\begin{tabular}{llcccccr}
\tablestrut&Atom&Present&CH&GU&LDA&PBE&\tablestrut\\ \hline
\tablestrut&He&0.01&-0.01&0.006  &0.07&0.01&\tablestrut\\
\tablestrut&Be&-0.07&-0.1&-0.02  &0.2&0.04&\tablestrut\\
\tablestrut&Ne&0.1&-0.07&0.05  &0.7&0.07&\tablestrut\\
\tablestrut&Be$^{2+}$&-0.02&-0.005&0.004  &0.2&0.04&\tablestrut\\
\tablestrut&C$^{4+}$&-0.02&-0.003&0.003  &0.4&0.06&\tablestrut\\
\tablestrut&C$^{2+}$&0.1&-0.2&0.01  &0.4&0.07&\tablestrut\\
\tablestrut&O$^{4+}$&0.1&-0.2&0.02  &0.6&0.1&\tablestrut\\
\end{tabular}
\smallskip
\caption{Error in the total energies of atoms for the new functional,
in Hartrees. For comparison, results are included for the CH and
the GU functionals and for the LDA and PBE-GGA approximations
within density functional theory.}
\label{table:atoms}
\end{table}

\end{document}